\documentclass[desactivate]{aa}  
\usepackage{graphicx}
\usepackage{xcolor}
\usepackage{txfonts}
\usepackage{hyperref}

\begin{document}

   \title{The selection function of the \emph{Gaia} DR3 open cluster census}

   \author{Emily L. Hunt
          \inst{1,2}
          \and
          Tristan Cantat-Gaudin\inst{2}
          \and
          Friedrich Anders\inst{3,4,5}
          \and
          Sagar Malhotra\inst{3,4,5}
          \and
          Lorenzo Spina\inst{6,7}
          \and
          Alfred Castro-Ginard\inst{3,4,5}
          \and
          Lorenzo Cavallo\inst{8}
          }

   \institute{
            Department of Astrophysics, University of Vienna, Türkenschanzstrasse 17, 1180 Wien, Austria\\
            \email{emily.lauren.hunt@univie.ac.at}
        \and 
            Max-Planck-Institut f\"ur Astronomie, K\"onigstuhl 17, 69117 Heidelberg, Germany
        \and 
            Departament de Física Quàntica i Astrofísica (FQA), Universitat de Barcelona (UB), Martí i Franquès, 1, 08028 Barcelona, Spain
        \and 
            Institut de Ciències del Cosmos (ICCUB), Universitat de Barcelona (UB), Martí i Franquès, 1, 08028 Barcelona, Spain
        \and 
            Institut d'Estudis Espacials de Catalunya (IEEC), Edifici RDIT, Campus UPC, 08860 Castelldefels (Barcelona), Spain
        \and 
            INAF, Osservatorio Astrofisico di Arcetri, Largo E. Fermi 5, 50125 Firenze, Italy
        \and 
            INAF, Osservatorio Astronomico di Padova, Vicolo dell’Osservatorio 5, 35122 Padova, Italy
        \and 
            Dipartimento di Fisica e Astronomia, Università di Padova, Vicolo dell’Osservatorio 3, 35122 Padova, Italy
        }

   \date{Received 21 October 2025; accepted 9 January 2026}

 
  \abstract
   {Open clusters are among the most useful and widespread tracers of Galactic structure. The completeness of the Galactic open cluster census, however, remains poorly understood.}
   {For the first time ever, we aim to establish the selection function of an entire open cluster census, publishing our results as an open-source Python package for use by the community. Our work is valid for the Hunt \& Reffert catalogue of clusters in \emph{Gaia} DR3.}
   {We developed and open sourced our cluster simulator from our first work. Then, we performed 80\,590 injection and retrievals of simulated open clusters to test the Hunt \& Reffert catalogue's sensitivity. We fitted a logistic model of cluster detectability that only depends on a cluster's number of stars, median parallax error, \emph{Gaia} data density, and a user-specified significance threshold.}
   {We find that our simple model accurately predicts cluster detectability, with a 94.53\% accuracy on our training data that is comparable to a machine-learning-based model with orders of magnitude more parameters. Our model itself offers numerous insights into why certain clusters are detected. We briefly used our model to show that cluster detectability depends on non-intuitive parameters, such as a cluster's proper motion, and we show that even a modest 25~km/s boost to a cluster's orbital speed can result in an almost 3$\times$ higher detection probability, depending on its position. In addition, we published our raw cluster injection and retrievals and cluster memberships, which could be used for a number of other science cases -- such as estimating cluster-membership incompleteness.}
   {Using our results, selection effect-corrected studies are now possible with the open cluster census. Our work will enable a number of brand new types of study, such as detailed comparisons between the Milky Way's cluster census and recent extragalactic cluster samples.}

   \keywords{open clusters and associations: general -- Methods: data analysis -- Galaxy: disc -- Galaxy: evolution 
               }

   \maketitle

\section{Introduction}\label{sec:introduction}

As the only galaxy that we observe from within, the Milky Way is a uniquely challenging galaxy to study. Nevertheless, our unique place in the Milky Way also allows for significantly more detailed observations of its content. To accurately untangle our Galaxy's structure from the Solar System, tracers with accurate distance information are essential. Open clusters (OCs) are among the most prolific such tracers. As single populations of stars with ages usually no higher than 1~Gyr \citep{Wielen1971, Pandey1986, Lamers2006, Anders2021},
the parameters of OCs can be inferred with much greater precision than for individual field stars -- either by averaging over all $N$ stars in a cluster to derive a measurement around $\sqrt{N}$ times more accurate, or by using a cluster's colour-magnitude diagram (CMD) to fit isochrones and derive a precise age or extinction value \citep{DemarqueLarson_1964,KharchenkoPiskunov_2013,Cantat-GaudinCasamiquela_2024}.

In the era of \emph{Gaia} \citep{GaiaCollaborationPrusti_2016}, the potential of OCs as Galactic tracers has only increased. The billion-star astrometric and photometric data from \emph{Gaia}'s second and third data releases \citep{GaiaCollaborationBrown_2018, GaiaCollaborationVallenari_2023} has enabled the discovery of thousands of new OCs since the release of \emph{Gaia} DR2 \citep[e.g.][]{Castro-GinardJordi_2018,Castro-GinardJordi_2019,Castro-GinardJordi_2020,Castro-GinardJordi_2022,SimLee_2019,LiuPang_2019,HuntReffert_2021,HuntReffert_2023}. At the same time, the cluster census itself has improved in quality, with over 1000 OCs from before \emph{Gaia} having been ruled out as asterisms \citep[][hereafter HR23]{Cantat-GaudinAnders_2020,HuntReffert_2023}. Despite these significant improvements
in the size of the OC census, modern OC catalogues are still highly incomplete. Extrapolating from the roughly complete solar neighbourhood, the Milky Way probably contains around $10^{5}$ OCs \citep{DiasAlessi_2002,Bonatto2006,KharchenkoPiskunov_2013,HuntReffert_2024}, and the latest OC catalogues contain barely 5\% of this total number of clusters \citep[e.g.][hereafter HR24]{HuntReffert_2024}, as extinction, crowding, and distance present firm limits on OC detectability. 
Studies that wish to infer some property of the Milky Way based on its observed population of OCs must hence face working with a largely incomplete cluster sample.
Traditionally,
the completeness issue was avoided by considering a volume-limited complete sample of clusters, usually by looking at the volume distribution of clusters and setting the completeness-limited radius to the point at which this distribution begins to fall. However, this self-driven method is inherently limited for two reasons: firstly, it is `data-wasteful', as the majority of a catalogue (outside of the 100\% completeness radius) must be discarded; and secondly, as it is insensitive to unknown incompletenesses or structure in a given dataset, this method is likely to produce incorrect results \citep{RixHogg_2021}. A more statistically rigorous approach is to derive the selection function of a given catalogue, which estimates the probability that a cluster with parameters, $\theta,$ will be detected by a given survey, $P(\text{detected}\,|\,\theta).$

Before the advent of {\it Gaia}, cluster catalogues compiled results from many different inhomogeneous surveys into a single work -- meaning that they would have a complicated selection function composed of each individual included survey. Modern machine-learning techniques and clustering algorithms, which are now commonplace for recovering OCs \citep[e.g.][]{Castro-GinardJordi_2018,HuntReffert_2023}, present an advantage: a single technique can detect a majority of the cluster population and create a single, homogeneous catalogue, which in turn will have a single selection function that could be determined.

In this work, we built upon our previous prototype work in the Galactic anti-centre \citep[][hereafter Paper I]{HuntCantat-Gaudin_2025} and constructed a selection function for the entire HR24 cluster catalogue using an injection-and-retrieval-based method. To do so, we simulated injection and retrieval following the clustering method of HR23, from which the HR24 catalogue was constructed. We show how seemingly innocuous facets of the OC census (such as its velocity distribution) can be severely impacted by not considering incompleteness. We published an open-source Python implementation of our model for use by the community. We outline our approach and our cluster simulator in Sect.~\ref{sec:methods}. We describe how we derived a selection function that depends on the number of stars in a cluster in Sect.~\ref{sec:models}. We provide a fast additional way to simulate the number of stars a cluster would have based on its parameters in Sect.~\ref{sec:nstar-predictor}. We show an example application of our selection function to demonstrate the impact of velocity on cluster detectability in Sect.~\ref{sec:uses}. Section~\ref{sec:caveats} discusses important caveats, best practices, and areas for future improvement of our model, and we conclude in Sect.~\ref{sec:conclusions}. For particularly keen readers, the appendices of this work include extended information, such as a number of example Python snippets for querying our model or simulating clusters and advice for future cluster-selection functions and cluster catalogues to make selection functions easier to determine.

\section{Cluster injection and retrieval strategy}\label{sec:methods}

In this section, we outline how we simulated, injected, and retrieved OCs to replicate the HR23 cluster recovery pipeline and derive its selection function. We begin with a brief recap of our Paper I cluster-simulation technique, before discussing additional features added to it to improve the legacy value of our published data of simulated and retrieved clusters. We then outline the parameter ranges explored in this work.

\subsection{Recap of Paper I cluster-simulation pipeline}\label{sec:methods:recap}

In Paper I, we created a cluster-simulation pipeline to simulate OCs across a wide range of parameters. We simulated OCs based on the assumptions and models recapped here; we refer the reader to Paper I for a more thorough explanation.

Our cluster simulation process begins by deriving absolute stellar photometry for a given cluster based on its age, $\log_{10} t$, and metallicity, $[\text{M}/\text{H}]$. We did this by interpolating PARSEC v1.2s isochrones \citep{BressanMarigo_2012} at a given $\log_{10} t$ and $[\text{M}/\text{H}]$ and then sampling stars of given masses from the cluster's isochrone based on a Kroupa initial mass function (IMF) \citep{Kroupa_2001} using the \texttt{imf} Python package;\footnote{\url{https://github.com/keflavich/imf}} this was done up to a total cluster mass of $M$. Assuming a single, fixed IMF for all of our clusters is a large but fair assumption for our purpose, as recent results have shown that many clusters still appear to retain the imprint of the IMF they were formed with (HR24), particularly for high-mass stars (which represent the majority of observed stars at the distances investigated in this study).

Having created a single-age, single-metallicity stellar population, the next step was to simulate the spatial and velocity information of each cluster-member star. We sampled positions and velocities for each cluster-member star around a mean cluster position and velocity in Cartesian space. For spatial positions, we assumed that clusters follow a \cite{King_1962} spatial profile with a core radius of $r_{c}$ and a tidal radius of $r_{t}$, also solving for the value of $r_{50}$ for each King profile following the method in HR23. Recent \emph{Gaia}-based results have shown that the bound cores of many clusters are surrounded by tidal tails and coronae \citep{RoserSchilbach_2019,MeingastAlves_2019,MeingastAlves_2021,TarricqSoubiran_2022,Kos_2024}, which are not included by King profiles; however, given the sparsity of these structures, we assume that their impact on cluster detectability is negligible, and we do not include them in our cluster simulations. To account for binarity, we pair stars into multiple systems based on the relations in \cite{MoeDiStefano_2017}, with secondaries placed near to their primary star based on their computed separation and a random orbital inclination. 

After sampling spatial information, we then sampled 3D stellar velocities from a Gaussian distribution \citep[which is a fair assumption for a star cluster][]{King_1966}, with a mean value given by the cluster's overall velocity and with a dispersion based on the cluster's mass \citep[][hereafter PZ10, their Eq.~4]{PortegiesZwartMcMillan_2010}. We assumed that $\eta=10$ in this equation, and then we linearly scaled the dispersion based on the assumed virial ratio of a cluster, $Q$ (where $Q=0.5$ is perfectly virialised). Throughout the simulation process, geometric calculations were performed by \texttt{astropy}'s coordinates module \citep{AstropyCollaborationRobitaille_2013,AstropyCollaborationPrice-Whelan_2018,AstropyCollaborationPrice-Whelan_2022}, which allows a cluster's member stars to be placed at a given Cartesian galactocentric position with given stellar motions, which can later be transformed into observed positions and velocities (i.e. spherical coordinates, parallaxes, and proper motions in \emph{Gaia}'s coordinate frame). We used the current galactocentric frame defaults in \texttt{astropy;} namely, the Sun at $X=-8122.0$~pc and $Y=0$~pc \citep{GRAVITYCollaborationAbuter_2018}, with a height above the plane $Z=20.8$~pc \citep{BennettBovy_2019}, and a galactocentric velocity of $V_X,V_Y,V_Z=(12.9,245.6,7.78)$~km/s \citep{DrimmelPoggio_2018}.

Then, this idealised simulated cluster was turned into a simulated cluster observation, which begins with a process of creating photometry and astrometry that mimic what \emph{Gaia} would observe for a given cluster. Firstly, using \texttt{astropy}'s coordinates module, observed positions and velocities were converted into Galactic coordinates $l$ and $b$, distance $d$, and proper motions $\mu_{\alpha^*}$ and $\mu_\delta$. Next, observed stellar photometric values were computed based on the cluster's distance and selected extinction, $A_V$, utilising the pre-computed \emph{Gaia} DR3 photometric relations in \cite{AndersKhalatyan_2022}. Then, the apparent magnitudes of cluster member stars were used to select similar, real \emph{Gaia} stars in a region around the cluster, assigning a real `twin' star to each cluster member. The astrometric and photometric errors of its real twin are assigned to simulated cluster-member stars. We created simulations of real measurements by re-sampling simulated observed astrometry and photometry from Gaussian distributions based on these error values.

We then proceeded to create unresolved binary stars in our simulated cluster observations. We converted simulated binaries into unresolved systems with summed photometry when their separation is below 0.6", which is a basic but fair resolution for \emph{Gaia} based on comparisons of \emph{Gaia} data to external binary star surveys \citep{FabriciusLuri_2021}. Notably, for reasons of computational efficiency, we did not simulate the influence of binary stars on observed proper motions and parallaxes -- despite the fact that binaries in \emph{Gaia} DR3 can have large errors in their astrometric parameters \citep{PenoyreBelokurov_2020,ChulkovMalkov_2022}, which can result in binary stars being missed as member stars of clusters \citep[e.g.][]{TagaevSeleznev_2025}. Depending on the period of the binary, this can manifest as a larger than normal re-normalised unit weight error (RUWE) value for the star. In practice, the impact of binarity on whether or not a member star is missed depends on period, distance, and orbital inclination of the system; only certain periods of binary star (such as those of around a few years) will have poor DR3 astrometry, and this effect is maximised for the nearest stars with the largest on-sky deviations from single-star astrometry. We expect the most significant errors for stars within a few hundred~pc of the Sun \citep{Castro-GinardPenoyre_2024} -- whereas this study only considers cluster detectability beyond 500~pc. For a cluster at a distance of 1~kpc with a 40\% binary-star fraction, extrapolating from simulations of binary-star detectability conducted by \cite{Castro-GinardPenoyre_2024} suggests that only around 2\% of cluster-member stars would be binaries with a high RUWE, which decreases even further as distance increases. Some additional binaries without a high RUWE (such as those with a period near one year) would also have poor astrometry. Not simulating binary-star astrometry hence contributes a systematic error to our results of the order of a few percent, albeit with the benefit of dramatically improved cluster-simulation speed.

At this point, our simulated clusters contain all cluster-member stars down to $G=21$, which is unrealistic given that \emph{Gaia} data become incomplete at the faint end \citep{FabriciusLuri_2021}. An important final step in our simulation process is to account for selection effects on cluster members. We modeled selection effects due to \emph{Gaia} data using the empirical selection function of \cite{Cantat-GaudinFouesneau_2023}, which models incompleteness at the faint end of \emph{Gaia} DR3 through comparison to external surveys. In addition, HR23's sample of \emph{Gaia} data relied on a number of quality cuts that also introduced membership-selection effects; they only performed clustering on stars with all five astrometric parameters, $G$, $G_{BP}$, and $G_{RP}$ photometry, and those with a \cite{RybizkiGreen_2022} v1 quality classification of at least 0.5. We modeled this selection effect using the method outlined in \cite{Castro-GinardBrown_2023}, which adds further (realistic) incompleteness at the faint end of our simulated clusters' membership lists, particularly in crowded regions. This concludes the simulation of a cluster.

\subsection{Additional features for this work}\label{sec:methods:additional}

\begin{figure*}[t]
\centering
\includegraphics[width=1.0\textwidth]{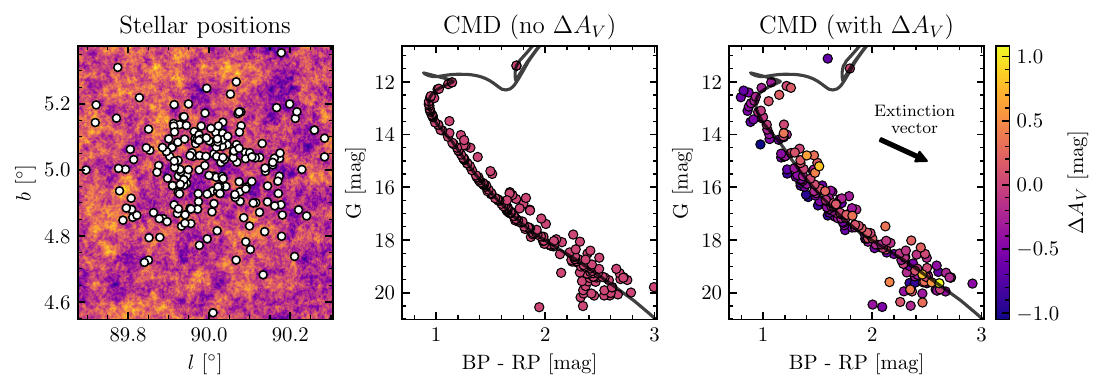}
\caption{Example of simulated cluster with photometry broadened by differential reddening. The cluster has an age of 1~Gyr, a mass of 500~M\textsubscript{\sun}, an extinction $(A_V)$ of 1.5~mag, and 0.3~mag of differential reddening (measured in colour: $E(B-V)$); it is at a distance of 1~kpc.
\emph{Left:} Random fractal noise map used to pick a differential reddening for each star. Member stars are shown as white circles, with the coloured background corresponding to changes in extinction, $A_V$.
\emph{Centre:} Simulated CMD for cluster without differential reddening, but including the impact of unresolved binary stars and \emph{Gaia} photometric errors. The black line shows the isochrone that the cluster's member stars were originally drawn from.
\emph{Right:} Final simulated CMD including impact of differential reddening. Member stars are colour-coded based on their differential reddening with the same colour scheme as the left panel. The labelled black arrow shows the mean extinction vector for stars in this cluster. For this cluster, differential reddening is strong enough to smooth out the clean sequence of unresolved binaries: main sequence stars with higher reddening move toward the binary sequence, and vice versa for binaries with lower reddening. \label{fig:methods_differential_reddening}}
\end{figure*}

Two notable improvements were made to our cluster simulation code for the purposes of this work, when compared to Paper I. Firstly, we added support for differential reddening to our simulated clusters; secondly, we open sourced our cluster-simulation code in the upcoming Python package \texttt{ocelot}\footnote{\url{https://github.com/emilyhunt/ocelot}}, helping to make our results reproducible.

As dust in our Galaxy has a complicated structure \citep[see e.g.][]{EdenhoferZucker_2024}, it is common for reddened OCs to have a different amount of extinction in different parts of the cluster, depending on the integrated structure of foreground dust clouds along the line of sight towards each member star. This causes a cluster's CMD to be broadened from the narrow, ideal isochronal line of stellar models. Typically, differential reddening is simulated by applying a random Gaussian-distributed reddening offset to each member star, depending on a given amount of differential reddening dispersion \citep[e.g.][]{HosekLu_2020}, motivated by the generally Gaussian dispersion of reddening values in clusters that have determined per-star extinctions \citep[e.g.][]{PancinoZocchi_2024}. However, differential reddening is spatially correlated across real clusters, with stars in similar positions having similar levels of reddening. This can lead to complicated changes in the shape of clusters, as certain parts of a cluster are reddened out of view \citep[][Fig. C.1]{Cantat-GaudinAnders_2020}, which is not otherwise simulated by non-spatial differential reddening. 3D dust maps do not have the required resolution to allow for real sub-parsec-scale variations in dust to be queried either, especially at the high end of distances considered in this study.
Instead, we approximated differential reddening by creating a spatially correlated random-noise map for each cluster. We did so with the common signal-processing concept of fractal noise \citep[also known as $1/f,$ or `pink' noise;][]{PhysRevLett.59.381}, a type of noise where power density is inversely proportional to frequency. This noise can be easily and quickly generated, and when used to select differential reddening values, the resulting dispersion in per-star differential reddening remains Gaussian -- except now it is spatially correlated. An example of a cluster's differential reddening noise map and the resulting cluster CMD is shown in Fig.~\ref{fig:methods_differential_reddening}. As a final motivation for this technique, we note that the chaotic and `cloudy' structure of fractal noise appears similar to the small-scale structure of high-resolution 3D dust maps \citep{EdenhoferZucker_2024}.

\subsection{Injection and retrievals}\label{sec:methods:injection}

\begin{table}[t]
    \caption{\label{tab:sampled_parameters}Range of cluster parameters explored in this study.}
    \centering
    \begin{tabular}{lp{2cm}ccc}
        \hline\hline
        Parameter & Source & Min. & Max. & Unit \\
        \hline
        
        $X$\tablefootmark{a} & $\mathcal{U}$ & -23.1 & 0 & kpc \\
        $Y$\tablefootmark{a} & $\mathcal{U}$ & -15 & 15 & kpc \\
        $Z$\tablefootmark{a} & $0.8\,\mathcal{N}(0,0.25)$ $+0.2\,\mathcal{U}$ & -2.5 & 2.5 & kpc \\

        $V_R$ & $0.8\,\mathcal{N}(0,35)$ $+0.2\,\mathcal{U} $ & -150 & 150 & km\,s\textsuperscript{-1} \\
        $V_{\phi-\text{circ}}$ &  $0.8\,\mathcal{N}(0,25)$ $+0.2\,\mathcal{U} $ & -150 & 150 & km\,s\textsuperscript{-1} \\
        $V_Z$ &  $0.8\,\mathcal{N}(0,15)$ $+0.2\,\mathcal{U} $ & -100 & 100 & km\,s\textsuperscript{-1} \\

        $M$ & Power laws\tablefootmark{b} & 40 & 50000 & M\textsubscript{\sun} \\
        $\log t$ & $\mathcal{U}$ & 6 & 10 & dex \\
        $[\text{M}/\text{H}]$ & $\mathcal{U}$ & -1 & 0.5 & dex \\
        $A_V$ & dust maps & 0 & 20 & mag \\
        $\Delta A_V / A_V$ & $\mathcal{U}$ & 0 & 0.5 & mag \\
        $Q$ & $\mathcal{N}(0.5, 4)$ & 0.1 & 10 & -- \\

        $r_c$ & $\mathcal{U}$ & 0.1 & 8 & pc \\
        $r_J$ & gal. potential & -- & -- & pc \\
        $r_t/r_J$ & $\mathcal{U}$ & 0.5 & 2 & -- \\

        \hline
    \end{tabular}
    \tablefoot{$\mathcal{U}$ denotes a uniform distribution between the minimum and maximum values. $\mathcal{N}(a,b)$ denotes a normal distribution truncated between the minimum and maximum values, with the mean, $a,$ and standard deviation, $b$. For some parameters, a weighted combination of both distributions was used.
    \tablefoottext{a}{Samples with a total heliocentric distance $d>15$~kpc or $d<0.5$~kpc were discarded.}
    \tablefoottext{b}{Samples from a broken power law that favoured low-mass clusters and is flat at high masses; that is, one with slopes (in mass order) of -1.013, -2.606, and 0.0, with break points at 813~M\textsubscript{\sun} and 3364~M\textsubscript{\sun}.}
    }
\end{table}

With our cluster simulation pipeline in hand, the last step in our method was to perform a collection of injection and retrieval experiments. We followed the clustering method of Paper I, which in turn replicated the clustering method of HR23 as closely as possible. In short, HR23 applied the Hierarchical Density-Based Spatial Clustering of Applications with Noise \citep[HDBSCAN;][]{HDBSCAN_original_paper,McInnes2017} clustering algorithm to discrete patches of \emph{Gaia} data to detect candidate clusters. HDBSCAN is a hierarchical, density-based clustering algorithm parameterised only on a cluster's minimum size, $m_\text{clSize}$, and a minimum-number-of-samples smoothing parameter. HR23 used HDBSCAN because of its superior sensitivity compared to other methods \citep{HuntReffert_2021}. To tessellate the sky into patches to cluster, HR23 used the Hierarchical Equal Area isoLatitude Pixelisation \citep[hereafter HEALPix][]{2005ApJ...622..759G} scheme, which divides spherical data into equal areas and approximately quadrilateral-shaped pixels while minimising spherical distortions. HEALPix pixels come in various `levels', with level zero containing 12 large pixels that cover the entire sky. Each increasing level contains four times as many pixels. HR23 mostly relied on runs at HEALPix level five, which tessellates the sky into 12288 pixels.

With the available methods, for each simulated cluster, the HEALPix level five pixel (in Galactic coordinates) of \emph{Gaia} DR3 data containing said cluster was loaded, in addition to the neighbouring eight pixels around it -- creating a 3$\times$3 dataset of nine level-five pixels to analyse with HR23's cluster-detection pipeline. As in HR23, each dataset's positions and proper motions were then transformed to a bespoke spherical coordinate system centred on its central pixel, minimising coordinate distortions in position or proper-motion data. The positions, proper motions, and parallaxes of each dataset were re-scaled into an arbitrary clustering dataset, where each of the five dimensions of the data has a zero median and unit-interquartile range to help improve clustering \citep[a step taken by HR23 and various other works, e.g.][given that clustering algorithms do not understand units]{Castro-GinardJordi_2018}. As in HR23, HDBSCAN was run with its minimum-number-of-samples parameter fixed at 10, and with the minimum cluster size, $m_\text{clSize}$, at four different values: 10, 20, 40, and 80. Notably, unlike in Paper I, we chose to inject and retrieve just one cluster at a time into each clustering field, ensuring that simulated clusters can never overlap within clustering fields and preventing `self-contamination' of our data -- an issue more prevalent at the lower distances considered in this study than in Paper~I. 

In HR23, they also included clustering runs at HEALPix level two and in Cartesian coordinates aimed at better detecting the full extent of nearby clusters, as some close clusters (such as the Hyades) are significantly larger than a single HEALPix level-five pixel. In this work, we only performed injection and retrievals at HEALPix level five, as HR24 is likely to be 100\% complete within the range where HEALPix level two or Cartesian clustering runs were used (see HR24, Sect.~5.1). Hence, the present work only injects clusters at a minimum distance of 500~pc for reasons of simplicity, which is the minimum valid distance of our selection function.

Having injected and retrieved clusters from \emph{Gaia} data, our next step following the HR23 pipeline was to recreate postprocessing steps they applied: namely, to compute the statistical significance of each cluster compared to their surrounding dataset, using the cluster significance test (CST) from \cite{HuntReffert_2021} -- a critical step that HR23 used in order to discard false positives. In their work, only clusters with $\text{CST}>3$ were retained, meaning that our selection function must establish both the probability of a cluster being detected by HDBSCAN and the probability of it having a CST greater than this threshold. HR23 recommends using a higher CST threshold of five or more when higher quality cluster samples are required, and so we also record detected cluster CSTs to investigate this further. Cluster CST values were used to select the best cluster detection given our four trialled $m_\text{clSize}$ values, prioritising the detection with the highest CST as in HR23.

The distributions from which the parameters of injected clusters were sampled are given in Table~\ref{tab:sampled_parameters}. Unlike in Paper I, we tried to sample wide ranges in parameters, including exploring if variations in parameters such as $[\text{M}/\text{H}]$ or $Q$ had an influence on cluster detectability, hence varying a total of thirteen parameters. For parameters such as cluster mass and cluster velocities, we tried to sample parameter ranges wider than the observed OC census, but with more sampling weight applied towards typical cluster values (such as by sampling from a combined Gaussian+uniform distribution).

Some additional cluster parameters were then computed based on the chosen parameters. For $A_V$, which is a function of cluster position, we again used the Bayestar19 3D dust map \citep{GreenSchlafly_2019} for most of our sample, in addition to using the relation $E(B-V)=0.884\times E(\text{Bayestar19})$ to convert the arbitrary reddening unit used by the Bayestar19 dust map into a $B-V$ reddening\footnote{See notes on Bayestar19 reddening unit conversion at \url{http://argonaut.skymaps.info/usage}}. To fill in the gaps not observed by the Pan-STARRS 1 survey on which Bayestar19 is based, we also used the \cite{ZuckerSaydjari_2025} dust map for an additional $\sim${}90$^\circ$ segment of the Galactic plane, as well as the dust maps of \cite{EdenhoferZucker_2024} and \cite{LallementVergely_2022} to help fill the pixels outside of the plane not covered by either of the first two maps. However, the latter two maps do not reach distances as high as the former two, meaning that the gaps in the Bayestar19 and Zucker map combinations cannot be completely filled with presently available data. For all maps, we assumed the dust curve shape $R_{V}=3.1$ -- although future studies may wish to incorporate 3D $R_{V}$ maps, such as that of \cite{ZhangGreen_2025}, which was released during the preparation of this work.

Our results also rely somewhat on a choice of Galactic potential. In Paper I, we fixed the cluster tidal radius, $r_t,$ to the cluster Jacobi radius, $r_J$, which gives the expected cluster size based on its mass and the local strength of the Galactic potential. In this work, we introduced a new parameter, $[r_t/r_J]$, a multiplier for $r_t$ that allows it to be smaller or larger than $r_J$, making our results less sensitive to a choice of Galactic potential. Nevertheless, we also opted to use a more recently derived potential for this work -- the MilkyWayPotential2022 in \texttt{gala} \citep[][hereafter MW22]{gala,gala_v_1_9_1}, which is based on the \cite{EilersHogg_2019} Galactic rotation curve and a range of recent Milky Way mass measurements. We also used the MW22 potential to compute the circular velocity of clusters $V_\text{circ}$ given their galactocentric radius, $R_{GC}$, which in turn was added to our sampled $V_{\phi-\text{circ}}$ velocities to compute their actual velocity in the $\phi$ direction, $V_\phi$. The final sampled cluster velocity was then converted to Cartesian coordinates (again using \texttt{astropy}'s coordinates module) and given to our cluster simulator to use as the mean cluster velocity when sampling stellar motions.

In total, we ran our full cluster simulation, injection, and retrieval pipeline on 80\,590 clusters across approximately one month of computational time on around 200 CPUs, using code that could run continuously until stopped. 4\,978 simulated clusters had zero observable stars in \emph{Gaia} DR3 and were hence recorded as non-detections without running clustering. Of the remaining injected clusters, 870 had membership lists that contained more stars from known clusters in HR23 than from their original simulated cluster. These overlapping cases were discarded, leaving a final dataset of 79,720 clusters. During the fitting of our models, we found that extinction values towards the Galactic centre for the Bayestar19 dust map (which covers $l>0^\circ$) and the \cite{ZuckerSaydjari_2025} dust map (which covers $l<360^\circ$) began to differ significantly at high distances ($d \gtrsim 3$~kpc), which could otherwise bias any selection function model working on either side of $l=0^\circ$ at high distances, which is an unfortunate side effect of the current lack of a single homogeneous dust map covering the full Galactic plane. Motivated by the fact that HR23 is virtually devoid of OCs in the Galactic centre due to the region's significant extinction and crowding, we did not use clusters with $X>-5$~kpc to fit our final selection function models, leaving a total of 62\,143 unbiased simulated cluster retrievals.

We publish our cluster injection and retrieval data along with this work, including the membership lists of simulated and retrieved clusters, as outlined in Appendix~\ref{app:published_tables} and available at the CDS. The remainder of the present paper focuses on establishing the selection function of the OC census, although our published data could also be used to model other aspects impacting cluster completeness, such as the completeness of cluster membership lists (see example in Appendix~\ref{app:membership_analysis}.)

\section{Training and validation of selection-function models}\label{sec:models}

\begin{figure}[t]
\centering
\includegraphics[width=\columnwidth]{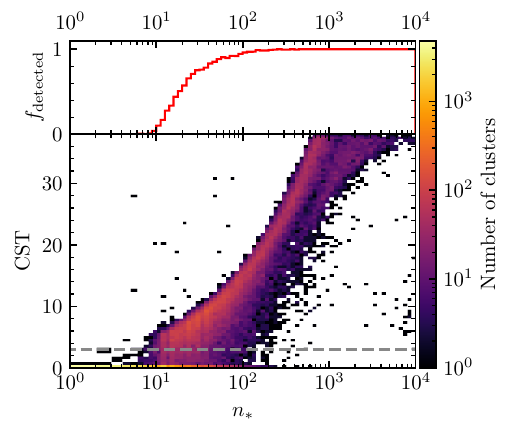}
\caption{\emph{Top:} Binned mean of $f_\text{detected}$ against number of simulated stars injected into \emph{Gaia} data for simulated clusters in this study. See also Fig.~\ref{fig:f_detected_trends} for more examples of the logistic dependence of $f_\text{detected}$ on $n_*$. \emph{Bottom:} 2D histogram of CST against number of injected stars for simulated clusters. The number of clusters at a given point is shown by the logarithmic colour-scale. The dashed grey line shows $\text{CST}=3$, which is the threshold for a cluster to be detected and included in HR23. Only clusters with a CST below 38 are shown; values larger than 38 are calculated as infinity by our Python method due to floating point precision. \label{fig:results_n_stars}}
\end{figure}

In this section, we outline the training and validation of our main selection-function model. We investigated a number of potential models and frameworks, aiming to create a method that could be easily replicated in future works for future cluster catalogues. 

In Paper I, we parameterised our cluster-selection function based on a number of cluster parameters, including their mass, distance, extinction, and age. For this work, we settled on a somewhat different, `split' approach; our selection function is instead parameterised on the true number of member stars of a cluster in \emph{Gaia} data, $n_*$, which we explore in this section. This provides two advantages: firstly, the number of stars visible in \emph{Gaia} data for a cluster is more interpretable, particularly when trying to understand the limitations of one cluster-detection method over another; second, when the number of stars in a cluster is already known (such as for a cluster that appears in one catalogue but not another), the selection function of one catalogue could easily be compared to another -- without requiring difficult-to-determine parameters such as a cluster's mass. 

In this section, we discuss how we created our $n_*$-based selection function, in addition to choosing additional parameters to add to our selection function to improve its accuracy, given that there is not a perfect correspondence between $n_*$ and how well a cluster is recovered by HR23's pipeline (Fig.~\ref{fig:results_n_stars}). When $n_*$ is not known, a cluster can be simulated based on its parameters to derive it (see Sect.~\ref{sec:methods}); we also provide a model to convert cluster parameters to $n_*$ in Sect.~\ref{sec:nstar-predictor}, for science cases where simulating clusters would be too time-consuming or impractical.

\subsection{Map of \emph{Gaia} data density}\label{sec:models:gaia_density}

\begin{figure}[t]
\centering
\includegraphics[width=\columnwidth]{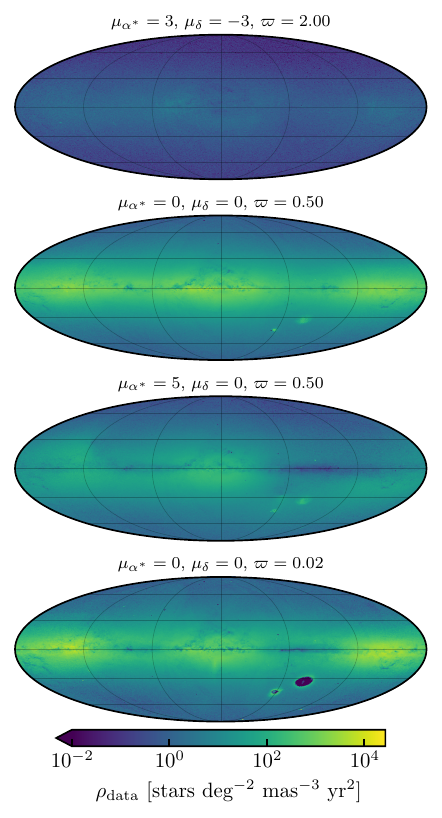}
\caption{Four maps of \emph{Gaia} data density at various proper-motion and parallax levels. The proper-motion and parallax values of each sub-plot are indicated in the titles, with proper motions given in milliarcseconds per year and parallax in milliarcseconds. Maps are plotted in a Mollweide projection in Galactic coordinates, with $l$ increasing to the left, north being up, and the Galactic centre in the middle. Vertical and horizontal grid lines show spacings of 60 and 30 degrees in $l$ and $b,$ respectively.\label{fig:density_map}}
\end{figure}

As in Paper I, parameters such as a cluster's extinction, age, distance, and orbit proved influential to whether or not a cluster is detected. To potentially reduce the dependence of our model on difficult-to-determine cluster parameters, we also experimented with including many other parameters, such as various measures of the density of \emph{Gaia} data around the cluster, $\rho_\text{data}$, the median magnitude of cluster member stars, or their median uncertainty in parallax $\text{med}(\sigma_\varpi)$ (which is a proxy for their overall scatter in the astrometric space in which clusters are detected).

To quickly determine the density of \emph{Gaia} data in a repeatable way, we measured the 5D astrometric density (i.e. density in position, proper motion, and parallax-space) of the HR23 sample of \emph{Gaia} data in a HEALPix level-seven map. Instead of measuring the density of \emph{Gaia} data at every possible proper motion and parallax, which would result in an extremely large and cumbersome map totalling tens or hundreds of GB in size, we instead only store the mean and covariance matrix of proper motions at given parallax values for stars in each HEALPix pixel. The density of HR23's sample of \emph{Gaia} data at an arbitrary position, proper motion, and parallax can then be estimated quickly as the value of a Gaussian probability density function based on these values. We measured the proper motion mean and covariance values in parallax bins of 0.1~mas in width from 0 to 2~mas, corresponding to the full distance range in our study. To improve resilience against stars with outlier proper motions, we also sigma clipped stars with proper motions more than 6$\sigma$ away from the mean proper motion in a given range. We linearly interpolated proper-motion mean and covariance values between parallax bins to ensure the smoothness of our map. This results in a map with 196\,608 pixels that each have 21 parallax bins, which can be efficiently queried using the Python package accompanying this work at a rate of around $10^5$ points per second.

Figure~\ref{fig:density_map} shows four example all-sky queries of this density map: one for a nearby parallax, two for a parallax of 0.5~mas, and one for a distant parallax corresponding roughly to the distance of the Magellanic Clouds. The maps show a clear increase in \emph{Gaia} data density with increasing parallax. Comparing the second and third maps shows one example of how changing proper motion causes asymmetric changes in \emph{Gaia} data density that depend on the fundamental distribution of stellar orbits in the Milky Way, with the proper motion of $\mu_{\alpha^*}$ of 5~mas~yr$^{-1}$ of the third map reducing density the most on the right of the Galactic centre. 

These maps highlight an important caveat of our present selection function. Our density method fails in the vicinity of existing `clusters' (in the dataset sense), which are shown, for instance, around the Magellanic Clouds (bottom right of fourth map in Fig.~\ref{fig:density_map}). In these cases, the cluster in the \emph{Gaia} data (i.e. the Magellanic Clouds) is more numerous than in field stars, causing it to dominate the derived density distribution. This could only be fixed by including a mixture model of density components in every HEALPix bin, which is beyond the scope of this work. Instead, we note that the selection function of whether a new cluster that is next to an existing cluster would be reported as a separate object or not in HR23 is too complicated to determine (Paper I); this is because cluster separation was a mostly manual process in HR23 that is impossible to replicate exactly. Moreover, it only impacts a small number of objects. The selection function of clusters in close proximity to one another is a known caveat of our work (Sect.~\ref{sec:caveats}) that should be addressed in the future, particularly through improved construction of cluster catalogues that could ease the determination of their selection function (Sect.~\ref{app:advice_for_future_catalogues}). In any case, none of the clusters in our training sample are close enough to the Magellanic Clouds for issues with our density measure in that region to impact them.

We queried $\rho_\text{data}$ for our injected and retrieved clusters with this map. Then, we proceeded to trialling different selection-function models.

\subsection{Selection of model parameters}\label{sec:models:xgboost}

\begin{figure}[t]
\centering
\includegraphics[width=\columnwidth]{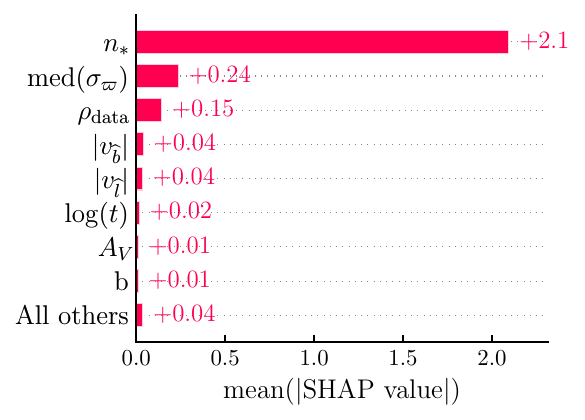}
\caption{Mean absolute SHAP values of XGBoost model used to investigate different selection-function parameters in Sect.~\ref{sec:models:xgboost}. \label{fig:xgboost_shap}}
\end{figure}

To find the other most important parameters governing the cluster selection function along with $n_*$, we used two complementary approaches. We began by making histograms showing cluster detectability as a function of cluster parameters to have an idea of which parameters may be important (Fig.~\ref{fig:app:detection_trends}). However, to estimate the importance of different parameters in a more qualitative and unbiased way, we trained basic machine-learning models and analysed which parameters appeared to be the most important for cluster detectability, using concepts from game theory. Our choice of model framework was XGBoost \citep{ChenGuestrin_2016}, which is a straightforward, fast, and flexible decision tree-based framework for inferring classification or regression relationships for tabular data, which sees common use in astronomy \citep[e.g.][]{AndraeRix_2023}. For all models, we report their accuracy after optimising their parameters with Optuna \citep{optuna_2019}, a model hyperparameter optimisation framework that can quickly converge on the most accurate and unbiased hyperparameter choices for a given method far more efficiently than human `random trials' of many different parameter combinations. This ensures that our reported tests are fair and use the optimal XGBoost parameters for each task.

To aid model fitting, we augmented our data in three ways. Firstly, our sample is divided into training and test datasets comprising 80\% and 20\% of our simulated clusters, respectively. Next, as clusters with zero stars in \emph{Gaia} data are never detected, and clusters with more than 1000 stars appear to always be detected, we trained our models only on clusters with $1 \leq n_* \leq 1000$, which improved inference speed with negligible loss of accuracy. Finally, we introduced an additional parameter, $\tau$, which is a user-defined CST threshold for a cut on the HR23 catalogue, and we chose a random value of $\tau$ between one and ten for each cluster in order to check its detectability. Hence, we allowed our models to work for both the default CST cut in HR23 ($\tau=3$) --for higher CST cuts for more high-quality samples-- and for lower CST cuts, to probe which clusters are only just outside the inclusion range in HR23.

Next, to qualitatively explain which aspects of our data contributed most strongly to cluster detectability, we used shapely additive explanation \citep[SHAP;][]{LundbergLee_2017} values to explain our XGBoost model's output. SHAP values apply concepts from game theory to machine-learning-model output, quantifying the strength by which certain features contribute to cluster detectability. We found that the median parallax uncertainty $\text{med}(\sigma_\varpi)$ and density of \emph{Gaia} data $\rho_\text{data}$ alone could explain almost all of the additional variance in cluster detectability, shown in Fig.~\ref{fig:xgboost_shap}. This is also demonstrated by comparing model binary classification accuracies (Table~\ref{tab:pdet_model_comparison}). Our best XGBoost model trained using just $n_*$ had an accuracy of 90.48\%. Including all cluster parameters and additional density or uncertainty measures almost halved the error rate of the model, corresponding to an accuracy of 94.68\%. Yet, a model trained just on $n_*$, $\text{med}(\sigma_\varpi)$, and $\rho_\text{data}$ saw almost no drop in classification accuracy despite requiring significantly fewer parameters, achieving 94.60\% accuracy. Hence, these three parameters (plus whichever $\tau$ value the user chooses) appear sufficient to achieve high model accuracy; we put remaining inaccuracy down to luck and whether or not our clustering algorithm happened to detect a given cluster.

\subsection{Markov chain Monte Carlo selection function}\label{sec:models:mcmc}

\begin{figure*}[t]
\centering
\includegraphics[width=\textwidth]{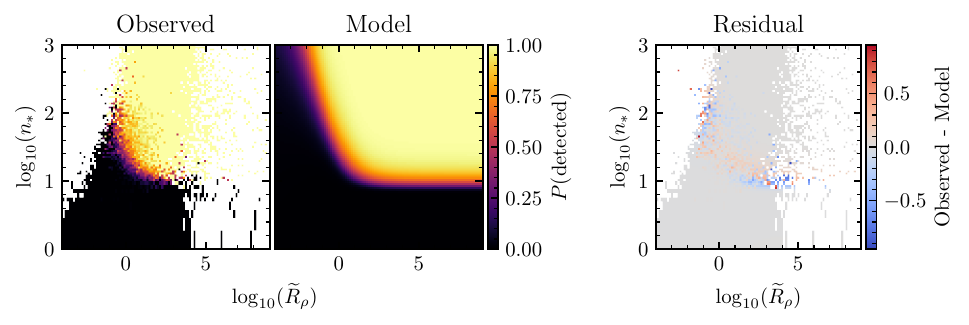}
\caption{Comparison between binned $P(\text{detected})$ for various $n_*$ versus $\tilde{R}_\rho$ values, along with predictions of our selection-function model, given $\tau=3$ (the CST threshold for detection of HR23). \emph{Left:} Observed distribution of injected and retrieved clusters. \emph{Centre left:} Predictions from our model. \emph{Right:} Remaining residuals between the model and observed data, where blue corresponds to model overconfidence, red to model underconfidence, and grey to correct model prediction. \label{fig:pdet_model_2d}}
\end{figure*}

\begin{figure}[t]
\centering
\includegraphics[width=\columnwidth]{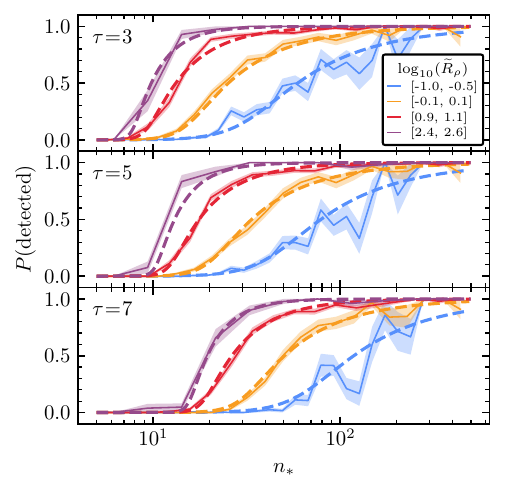}
\caption{Comparison between injection and retrieval data (solid line, shaded region shows uncertainty) and our model (dashed line) for various CST thresholds, $\tau,$ and restricted to different density ranges, as indicated in the figure legend. Plots from top to bottom show $\tau$ levels of 3, 5, and 7, respectively. \label{fig:pdet_model_comp}}
\end{figure}

\begin{table}[t]
    \caption{\label{tab:pdet_model_comparison}Comparison between XGBoost and MCMC models for predicting cluster detectability from number of stars.}
    \centering
    \begin{tabular}{llcc}
        \hline\hline
        Method & Parameters given & $\mathcal{A}$ (\%) & $\mathcal{L}$ \\
        \hline
XGBoost & $n_*,\tau$ & 90.48 & -2653 \\
XGBoost & $n_*,\tau,\text{med}(\sigma_\varpi),\rho_\text{Gaia}$ & 94.60 & -1693 \\
XGBoost & All\tablefootmark{a} & 94.68 & -1587 \\
MCMC & $n_*,\tau,\tilde{R}_\rho$ & 94.53 & -1672 \\
        \hline
    \end{tabular}
    \tablefoot{\tablefoottext{a}{All cluster parameters and \emph{Gaia} density parameters were given to the model, for a total of 19 parameters.}}
\end{table}

Noting that our selection function does not appear to require many parameters, and that the relationship between a cluster's detectability and $\log_{10}(n_*)$ appears to be a logistic (Fig.~\ref{fig:results_n_stars}, top panel), we fitted a simple logistic model to our results to create our final selection function -- which has the advantage of significantly improved interpretability over a machine-learning approach. Inspired by \cite{Cantat-GaudinFouesneau_2023}, which also fitted a logistic model to derive a selection function for \emph{Gaia} DR3, we fitted a generalised logistic model for the detection probability of a cluster P(\text{detected}) of the following form:

\begin{equation}
P(\text{detected})=1 - \left[ 1 + \exp\left(\frac{\log_{10}(n_*) - a}{\sigma}\right) \right]^{-k},
\end{equation}

\noindent
where $\sigma$ is the width of the logistic, $k$ is a skew parameter that must be greater than zero, and $a$ is an offset to $n_*$. $k=1$ corresponds to an unskewed logistic at which $a$ defines the location of the 50\% detection probability of a cluster. By default, $a$ and $k$ in this model are highly inter-dependent; to improve how well our model can be fitted and interpreted, we instead define $a$ in terms of a more interpretable parameter, $n_{*,50}$, which is the number of stars required for a cluster to have a 50\% detection probability given $k$ and $\sigma$. Rearranging for $a$ when $P(\text{detected})=0.5$ and $\log_{10}(n_*)=\log_{10}(n_{*,50})$, this means that $a$ is now given by

\begin{equation}
a = \log_{10}(n_{*,50}) - \sigma \log\left( 2^{1/k} - 1 \right).
\end{equation}

Given that $n_{*,50}$, $\sigma$, and $k$ could all be arbitrary functions that depend on $\tau$, $\text{med}(\sigma_\varpi)$, and/or $\rho_\text{data}$, choosing a correct functional form of each to create an accurate $P(\text{detected})$ equation is non-trivial. Initially, we experimented with using symbolic regression with PySR \citep{Cranmer_2023} to derive functional forms for $n_{*,50}$, $\sigma$, and $k$, given $x=\log_{10}(n_*)$. Although the complexity of the equations PySR derived to achieve our desired accuracy was unsatisfactory, PySR nevertheless provided new insights. 

The main insight we gained was the hypothesis that the reason why a cluster is or is not detectable in \emph{Gaia} data can be almost entirely explained by how dense it is compared to the data that surrounds it. Fundamentally, cluster detection in HR23 (or any other cluster-detection study with clustering algorithms) depends on how dense a cluster is in \emph{Gaia} data. The ratio of the density of a cluster to the density of \emph{Gaia} data at the location of the cluster, $R_\rho=\rho_\text{cluster} / \rho_\text{data}$, could then be the other main variable required in our selection function.

Defining the density of a cluster, $\rho_\text{cluster}$, is non-trivial, as star clusters have a smooth range of different densities. In position space, their \cite{King_1962} density profile smoothly drops from a maximum value to zero; likewise, in astrometric space, clusters have an approximately Gaussian velocity dispersion due to \emph{Gaia} measurement errors and their internal velocity dispersion. We opted for a single, representative value that is straightforward to calculate. Noting that errors on proper motions and parallaxes have a similar order of magnitude in \emph{Gaia} data, and that a cluster's astrometric dispersion is dominated almost entirely by this error, we define an approximate cluster density simply as

\begin{equation}
\rho_\text{cluster,approx.} = \frac{n_{*}}{\text{med}(\sigma_\varpi)^3},
\end{equation}

\noindent
which gives an approximate measure of cluster density in proper motion and parallax space. We also trialled using a measure of cluster size in positions, but this was generally unhelpful -- in agreement with our results in Paper I, where we found that the physical size of a cluster has negligible effect on cluster detectability. As $\rho_\text{data}$ is defined per degree, some correction for area is still necessary to derive physically meaningful ${R}_\rho$ values -- so we defined our density ratio as `per-field', dividing by the area of HR23 clustering fields (30.21$^\circ$) to express ${R}_\rho$ purely in terms of proper motion and parallax density. We hence defined a (approximate) cluster-to-\emph{Gaia} -density ratio as

\begin{equation}
\tilde{R}_\rho=\frac{\rho_\text{cluster,approx.}}{\rho_\text{data} / 30.21},
\end{equation}

\noindent
and we tested using it as a single variable in the functional form of $n_{*,50}$, $\sigma$, and $k$.

Plotting $n_*$ against $\tilde{R}_\rho$ reveals an intuitive relationship. The left panel of Fig.~\ref{fig:pdet_model_2d} shows that as $\tilde{R}_\rho$ decreases, $n_*$ needs to increase for a cluster to be detected: a less dense cluster requires more member stars to compensate for its lack of density through the quantity of visible members. However, this relationship is non-linear, and high density values reveal a minimum $n_*$ value below which clusters are only rarely recovered. This minimum value of around ten is likely related to HDBSCAN's minimum cluster size, which had a lowest trialled value of ten in HR23.

We trialled a number of different functional forms for $n_{*,50}$, $\sigma$, and $k$ that allowed them to vary as a function of $\tilde{R}_\rho$. We also investigated their dependence on the CST threshold, $\tau$, finding that this dependence was largely constrained to $n_{*,50}$. Hence, we fitted models of the following form:

\begin{gather}
\log_{10}(n_{*,50}) = g_{n_{*,50}}( \tilde{R}_\rho ) + u \tau ^v ,\\
\sigma = g_\sigma( \tilde{R}_\rho ) ,\\
k = g_k( \tilde{R}_\rho ),
\end{gather}

\noindent
where $u$ and $v$ are constants defining a power-law dependence of $n_{*,50}$ on $\tau$. To allow the three logistic parameters to transition from a single value at high $\tilde{R}_\rho$ to a linear relation at low values, we defined a smoothing function (similar to a `softplus' activation function in machine learning) given by

\begin{equation}
g_x( \tilde{R}_\rho ) = \frac{m_x}{|d_x|} \log \left( 1 + \exp \left(d_x (\log_{10}(\tilde{R}_\rho) - b_x) \right) \right) + c_x,
\end{equation}

\noindent
where $x$ is the variant of the smoothing function, $m_x$ is the gradient of the smoothing function in the increasing part, $c_x$ is a constant defining the minimum or maximum value of the function, $b_x$ is a horizontal offset, and $d_x$ determines how sharp the `knee' between the flat and increasing parts is, in addition to its orientation. This four-parameter function was found to capture differences between all parameters well. We trialled setting some of the four variables per smoothing function (such as offset, $b_x$ or knee sharpness, $d_x$) as identical for all three logistic parameters; however, leaving all four free to vary for all parameters was found to produce a model providing the most accuracy.

After setting some relatively uninformative priors on our 14 model parameters to ensure fitting efficiency and stability (see Table~\ref{tab:parameters}), we fitted our selection-function model to our data using \texttt{emcee} \citep{Foreman-MackeyHogg_2013}. \texttt{emcee} is a simple Markov chain Monte Carlo (MCMC) sampler that can be used to fit models and sample from their parameters, estimating both parameter values and their corresponding uncertainties. \texttt{emcee} requires an initial parameter guess, which we found by initially optimising our model's parameters with Optuna. To fit our model, we used a standard classification log loss, $\mathcal{L}$ \citep[see][]{Cantat-GaudinFouesneau_2023}. We performed 10\,000 MCMC steps, discarding the first 1000 as burn-ins. Our fitted parameter values and further plots of our model are available in a table and figures in Appendix \ref{app:model_fit_results}, or in the accompanying Python package to this paper that includes our code implementation of this model. Our final model has an overall 94.53\% classification accuracy on our training data, which is only marginally lower than the best XGBoost models we trialled, despite using fewer input parameters (Table~\ref{tab:pdet_model_comparison}) -- validating that our model parameterised only on $n_*$, $\tilde{R}_\rho$, and $\tau$ is highly accurate, even compared to machine-learning models with far more free parameters.

Comparing the distribution of $n_{*}$ and $\tilde{R}_\rho$ in Fig.~\ref{fig:pdet_model_2d} for our injected and retrieved clusters, our model creates a good prediction with minimal residuals, except around the edges of this distribution, which are less well sampled by our training data. A future iteration of this work could try to sample clusters more uniformly in $n_*$-$\tilde{R}_\rho$ space to ensure that training data samples this region well. 
Figure~\ref{fig:pdet_model_comp} shows that our logistic model 
closely matches the distribution of our recovered injections and retrievals for a range of different $\tau$ CST thresholds.

\section{XGBoost $n_*$ predictor}\label{sec:nstar-predictor}

\begin{figure}[t]
\centering
\includegraphics[width=\columnwidth]{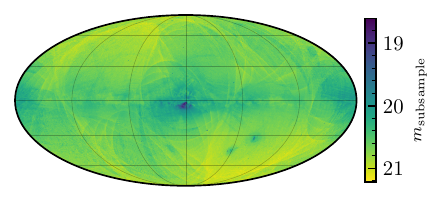}
\caption{Map of median $G$-band magnitude of stars removed by HR23's \emph{Gaia} data cuts, $m_\text{subsample}$. The map is plotted in the same projection as Fig.~\ref{fig:density_map}. \label{fig:m_subsample}}
\end{figure}

\begin{figure}[t]
\centering
\includegraphics[width=\columnwidth]{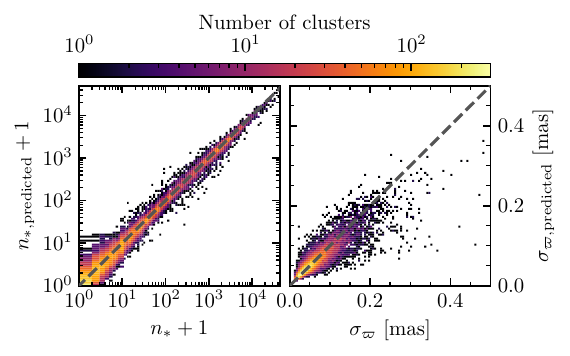}
\caption{Accuracy of XGBoost $n_*$ and $\sigma_\varpi$ predictor in Sect.~\ref{sec:nstar-predictor} on its test dataset, with true values on the $x$ axes and predicted ones on the $y$ axes. A perfect prediction falls on the dashed grey line.  \label{fig:n_stars_model_accuracy}}
\end{figure}

As an alternative to full cluster simulation to transform physical parameters such as mass and age into observed ones, $n_*$ and $\text{med}(\sigma_\varpi)$, we also created an XGBoost model to streamline the cluster simulation and prediction of $n_*$ and $\text{med}(\sigma_\varpi)$ for certain selection-function use cases. While our full cluster simulation suite was open-sourced in\texttt{ ocelot}\footnote{\url{https://github.com/emilyhunt/ocelot}} as a part of this work, it takes approximately one second (not including time to load \emph{Gaia} data) to simulate a cluster with it -- which could quickly become prohibitive for larger works. We therefore optimised the parameters of an XGBoost model tasked to predict $n_*$ and $\text{med}(\sigma_\varpi)$. To fit these models, we did not discard clusters at high $X$ as we did with the selection function -- giving us 79\,720 clusters, a training sample of 59\,790 clusters, and a test sample of 19\,930 clusters. We used Optuna to optimise model parameters. We set the number of estimators, \texttt{n\_estimators,} to be fixed at 100, as larger values dramatically increased training and inference time without significant accuracy improvements. As sampling of the IMF is stochastic, and a given set of cluster parameters does not produce an exact $n_*$ value, we also ensured that this method had quantified uncertainties by creating a bootstrap model ensemble of 250 models trained on independent samples of 66.6\% of our training data.

Initially, we trained models that included the cluster parameters $M$, $A_V$, $\Delta A_V$, $d$, $\log t$, and $[\text{M}/\text{H}]$. However, as selection effects are significant contributors to incompleteness at the low-mass end, and these effects are position dependent, we also incorporated selection-effect estimates. Firstly, we included the parameter $m_{10}$ from \cite{Cantat-GaudinFouesneau_2023}, which is the median magnitude of stars that do not appear in \emph{Gaia} data. In a similar vein, to quantify sub-sample-selection effects for our $n_*$ predictor, we also measured the median magnitude of stars that appear in \emph{Gaia} but are rejected from HR23's analysis due to their selection of \emph{Gaia} data, which we denote $m_{\text{subsample}}$. Our map of $m_{\text{subsample}}$ is shown in Fig.~\ref{fig:m_subsample}.

Finally, we found that changing XGBoost's loss function to be different to its default value produced a less biased model. By default, XGBoost uses a mean squared error, which penalises incorrect predictions based on their absolute size -- given the large $n_*$ range in this study (zero to over 50\,000), this generally forced the model to prioritise correct predictions at high $n_*$ values. To instead penalise predictions based on the order of magnitude of how incorrect they are at all $n_*$ values, we instead tasked XGBoost with predicting $\log_{10}(n_*+1)$ (which is conceptually the same as using the mean squared logarithmic error, but with the one added to $n_*$ to account for $\log_{10}(0)$ being undefined). This produced a model that was accurate across the whole range of $n_*$ and $\text{med}(\sigma_\varpi)$ (Fig.~\ref{fig:n_stars_model_accuracy}), which we release with this work.

\section{An example selection function use: The impact of a cluster's orbit on its detectability}\label{sec:uses}

\begin{figure}[t]
\centering
\includegraphics[width=\columnwidth]{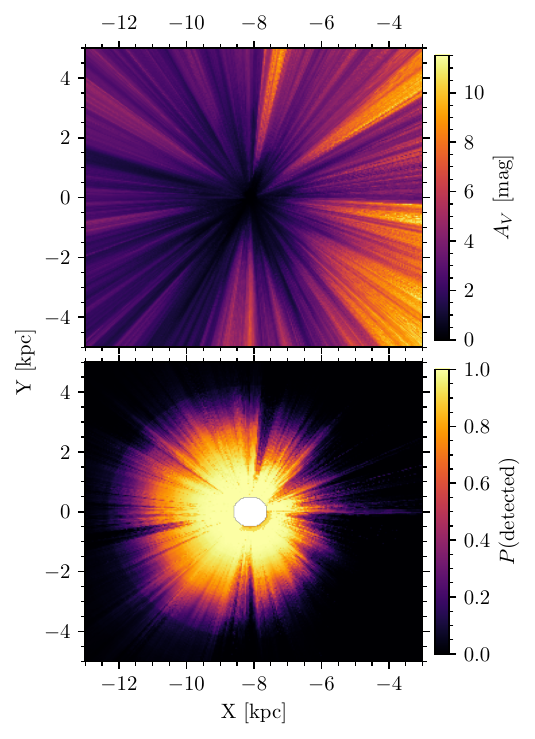}
\caption{Example of application of our full selection function pipeline to map cluster detectability. \emph{Top:} Adopted dust map viewed top-down in galactocentric Cartesian coordinates with the Galactic centre to the right, averaged from $z$ values from -100 to 100~pc. \emph{Bottom:} As above, but predicting the detection probability of a 2.5~Gyr old, 200~M\textsubscript{\sun} cluster with solar metallicity on a circular orbit. $n_*$ and $\text{med}(\sigma_\varpi)$ were predicted with our XGBoost predictor and then fed to our MCMC selection-function model. The white circle in the middle is a masked region, as distances lower than 500~pc were not considered in this study. \label{fig:detection_and_dust}}
\end{figure}

\begin{figure*}[t]
\centering
\includegraphics[width=\textwidth]{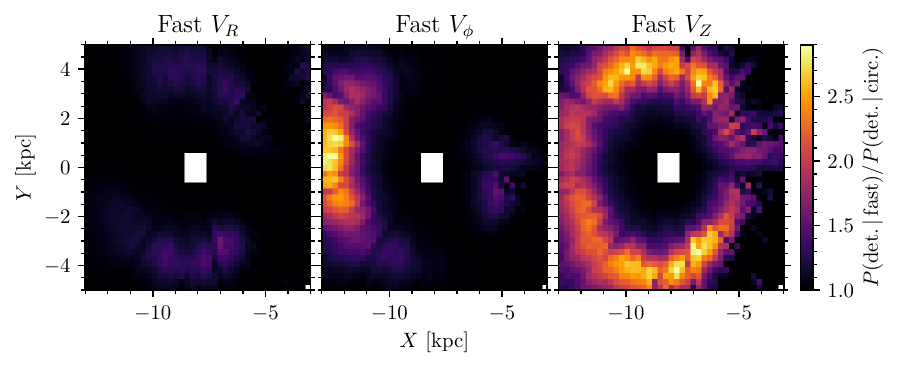}
\caption{Impact of cluster detectability from clusters being on non-circular orbits, shown for the same cluster as in Fig.~\ref{fig:detection_and_dust} and plotted in the same coordinate system. From left to right, plots show the factor by which cluster detection probability is increased by applying a 25 km/s boost to cluster orbital speed in the radial (\emph{left}), $\phi$ (\emph{centre}), and $Z$ (\emph{right}) directions, averaged from $z=-100$~pc to $z=100$~pc. The white region in the middle is for distances lower than 500~pc (which were not considered in this study).\label{fig:n_50_with_velocity}}
\end{figure*}

We plan to explore scientific uses of our selection function in future works. As an illustrative example of our selection function's utility, in this section, we demonstrate use of our full model pipeline, and briefly show how even modest increases to cluster orbital speed can dramatically increase cluster detectability. 

Figure~\ref{fig:detection_and_dust} shows the detectability of a 2.5~Gyr old 200~M\textsubscript{\sun} cluster with solar metallicity on a circular orbit, having predicted its number of stars and median parallax error with our XGBoost model within a 5~kpc box around the Sun. To construct the figure, we queried our selection function for circular orbits with orbital speeds 
taken from the MW22 potential, also using the Bayestar19 and \cite{ZuckerSaydjari_2025} dust maps to estimate extinction in 3D. The figure clearly shows how a cluster with such parameters is effectively completely undetectable to HR24 within around 4~kpc of the Sun. The `fingers-of-god' effect in Fig.~\ref{fig:detection_and_dust} is caused by foreground extinction, which causes greater difficulty in seeing clusters along certain lines of sight.

We show how the detectability of such a cluster can be boosted by non-circular orbits in Fig.~\ref{fig:n_50_with_velocity}, which demonstrates the increase in detectability from a 25~km/s boost to the cluster's orbital speed within a few kiloparsecs of the Sun, shown for a boost in each of the three galactocentric cylindrical coordinates. Even for a modest 25~km/s speed boost, which is well within the range of scatter in the observed cluster's velocity distribution \citep{TarricqSoubiran_2021}, cluster detectability is significantly boosted by faster-than-circular cluster orbits. Particularly in the $z$ (vertical) and $\phi$ (in direction of disc rotation) directions, this boost can cause the example cluster to be almost three times easier to detect. The magnitude of this boost is strongly direction and distance-dependent, with a complicated polar detectability pattern imprinted on the $\phi$ and $\rho$ directions, which are strongest when all of the cluster's 3D velocity boost goes in the direction of proper motions measured by \emph{Gaia} (radial velocities are not used by HR23 for clustering). In addition, the detectability increase of a boost likely also depends on background stellar motions: boosts to the $Z$ and $\phi$ directions are weakest towards the Galactic centre, probably because the Galactic bulge is both denser and has high scatter in stellar orbits. For clusters that are easier to detect (i.e. younger or more massive ones), this result and shape still hold; the ring of increased detectability is simply shifted to higher distances.

The impact of orbit on cluster detectability across the Galactic disc suggests that recent results using the OC census to trace the velocity distribution of OCs may need to reviewed. The complicated polar pattern of detectability boosts would cause the tails of the overall orbital velocity dispersions of OCs to appear wider than they actually are, as clusters with extreme orbits are more easily detected. Even worse, this polar pattern aligns somewhat with local spiral arms \citep{PoggioDrimmel_2021,Castro-GinardMcMillan_2021,ZuckerAlves_2023}. Particularly towards the anti-centre, a boost to $V_\phi$ can significantly increase cluster detectability; however, these regions also correspond to the location of a known spiral structure. Similarly, clusters moving quickly in the $\rho$ direction in the inter-arm region ahead of and behind the Sun's orbit will also receive a boost. Any study using the census of OCs to trace the velocity imprint of spiral arms would find that the arms away from and towards the Galactic centre have higher than expected dispersion in $V_\phi$; on the other hand, the Local Arm containing the Sun will appear to have more radial (towards the Galactic centre) dispersion ahead of and behind the Sun's orbit. A ring around the Sun will also have higher than expected dispersion in $V_Z$.

In general, in Galactic dynamics, it is common to assume that velocity distributions are easier to infer than spatial ones, as samples can usually be assumed not to have a detection bias as a function of their velocity \citep[e.g.][]{TarricqSoubiran_2021,AlfaroSanchez-Gil_2025}. While this assumption holds for stars, as their probability of inclusion in a dataset is entirely brightness dependent \citep{HuntVasiliev_2025}, more care must be taken for OCs in the \emph{Gaia} era: proper motions are an essential component for detecting clusters in \emph{Gaia}, but this means that fast-moving clusters are more likely to be detected than slow-moving ones, causing any simple measurement of the dispersion of the cluster census to be biased.

\section{Caveats and advice for effective use of our model}\label{sec:caveats}

Before using our selection function in future work, it is worth recapitulating and explicitly stating some of the caveats of our selection function. Firstly, and most obviously, it is only valid within the cluster parameter ranges that we explored in this work (Table~\ref{tab:sampled_parameters}); works focused on clusters with parameters outside of this range, or clusters that do not follow a King profile (such as dissolved or unbound clusters), are not guaranteed to receive correct predictions from our selection function.

A second caveat of our work is that we cannot replicate all aspects of the HR23 and HR24 cluster-catalogue pipelines. In particular, HR23 ran a large number of different clustering runs in different fields which had to be merged back together. Although this was a well-tested and mostly automated process, it is still possible that some clusters could have been lost in this process, or in other aspects of the HR23 cluster catalogue pipeline. We caution that our selection function is not a definitive answer to whether or not an individual cluster should appear in HR23, and errors could exist in a limited number of cases. We provide some recommendations for future cluster blind searches that would alleviate these issues in Appendix~\ref{app:advice_for_future_catalogues}.

A third caveat of our work is that it only predicts the probability that a real cluster appears in HR23 (i.e. the true positive rate), and not the probability that a false one appears there (i.e. the false-positive rate). In principle, the CST cut and CMD classifier from HR23 and object classifications from HR24 can aid with restricting their catalogue to an exclusively high-quality sample, highly likely to be real objects, and we refer the reader to HR23 and HR24 for discussions of best practices for using these parameters. Our selection function allows for the setting of different CST thresholds, $\tau$, and we recommend taking advantage of this for future studies -- for instance, by setting $\tau=5$ or $\tau=10$ and cutting HR24 based on that level (in addition to restricting it to objects classified as open clusters), and hence working only with a particularly high-quality sample of their work.\footnote{We did not establish the selection function of HR23's CMD classifier in this paper; their CMD classifications are an optional catalogue filter that can result in the removal of genuine clusters that have broadened CMDs from a young age or an embedded status \citep{QuintanaHunt_2025}.} If computationally possible, we recommend that works try running their analysis at a few different $\tau$ levels, explicitly reporting whether or not using different samples of HR24 causes a qualitative change to their results that could be a result of false positives emerging at low $\tau$ levels.

The final caveat of our work is that many aspects of using our selection function rely on accurate cluster parameters being available, which may often not be the case. In this work, we effectively reduced our selection function's dependence on cluster parameters by splitting it into a function that only depends on basic \emph{Gaia} observables ($n_*$, $\rho_\text{data}$, and $\text{med}(\sigma_\varpi)$), which can be used in cases when these parameters are already known, and separate tools for simulating clusters or predicting $n_*$ when a use case needs to work with more basic cluster parameters. This second use case is much more vulnerable to this caveat. For instance, 3D dust maps must be used to estimate the extinction towards a hypothetical cluster, and so any science case working in this way is only as good as currently available 3D dust maps -- all of which have limited distance and angular resolution, and none of which can confidently cover the entire Galactic plane out to more than a few kpc. Similarly, when using cluster parameters in HR24, it is important to appreciate that cluster parameters remain far from perfect and depend strongly on adopted methodology and cluster membership \citep{Cantat-GaudinCasamiquela_2024,CavalloSpina_2024}. 

Future studies improving cluster parameters and 3D dust maps further would help to alleviate this final caveat, and such studies should continue to be pursued -- particularly as the present paper sits on the horizon of many upcoming data releases that will greatly aid in 3D dust mapping and cluster-parameter determination. We encourage authors conducting parameter-determination studies to develop and apply their method to all clusters in HR24, and not just a sub-set of clusters, as this avoids the need to account for yet another selection function.

With \emph{Gaia} DR4 scheduled to be released in December 2026 at the time of writing\footnote{\url{https://www.cosmos.esa.int/web/gaia/release}}, it is likely that new, expanded cluster catalogues will become available soon. We hope to see our work repeated on future data releases and catalogues; in that vein, we also list some recommendations for future selection-function studies in Appendix~\ref{app:advice_for_future_sfs} that will aid in their construction.

\section{Conclusions}\label{sec:conclusions}

In this work, we established the first ever global selection function of the OC census, which is valid for the HR24 catalogue. Building on our work in Paper~I, we created an accurate cluster simulator and open sourced it in the \texttt{ocelot} Python package. We performed 80\,590 cluster injection and retrievals into \emph{Gaia} DR3, varying a wide range of different parameters in order to thoroughly investigate cluster detectability. Next, we created a fast way to look up \emph{Gaia} data density at a given location, proper motion, and parallax, and used it to create a cluster-selection function that only depends on a cluster's number of stars, \emph{Gaia} data density, median parallax error, and a chosen cluster-significance cut of the HR24 catalogue. In addition to our cluster simulator, we also provide a fast XGBoost model to predict a given cluster's number of stars and median parallax error based on its parameters.

In the spirit of open science, we have open sourced our cluster simulator and selection function for use of the community. Our selection function is available as a Python package\footnote{\url{https://github.com/emilyhunt/hr_selection_function}}. We also release the results of our injection and retrieval experiments, providing a large catalogue of simulated cluster injections and retrievals. These raw data can be used to extend our present work by investigating cluster membership list incompleteness and purity, for example (see example in Appendix~\ref{app:membership_analysis}). To aid use of our work, this paper includes an explicit-caveats section aimed at informing users of best practices when using our results.

Open clusters have been widely catalogued for hundreds of years, but a thorough understanding of exactly how many are contained in a present-day catalogue had so far been elusive. Our work will enable a number of brand new, selection-effect-corrected studies to be conducted. We briefly demonstrated one such utility of our selection function in this work, showing that the detectability of OCs is strongly impacted by their proper motion (i.e. speed and direction of Galactic orbit) -- showing that previous results utilising cluster velocities directly will have derived biased properties about the cluster census. Our work will also enable better comparisons between the Milky Way and other galaxies, which is timely given recent strides in the cataloguing of extragalactic clusters such as through the PHANGS survey \citep{MaschmannLee_2024}. Parameters such as the cluster age and mass function \citep[e.g.][]{Anders2021,HuntReffert_2024,AlmeidaMoitinho_2025} could now be derived as a function of Galactocentric radius, and compared to other galaxies.

The fundamental objective of this short series was to investigate if the selection function of the OC census is computationally possible to derive. With this work, we hope we have demonstrated that concretely. We encourage future authors of cluster catalogues to use our methodology and to continue to derive OC-catalogue selection functions for their catalogues based on upcoming surveys and data releases.

\section*{Data availability}

Tables~\ref{tab:clusters} and \ref{tab:members} are only available in electronic form at the CDS via anonymous ftp to cdsarc.cds.unistra.fr (130.79.128.5) or via https://cdsarc.cds.unistra.fr/cgi-bin/qcat?J/A+A/. Our models are available as a Python package at \url{https://github.com/emilyhunt/hr_selection_function}.

\begin{acknowledgements}

We thank the anonymous referee for their helpful comments on our manuscript, which especially improved its clarity. ELH thanks Hans-Walter Rix for helpful discussions. This work has made use of data from the European Space Agency (ESA) mission {\it Gaia} (\url{http://www.cosmos.esa.int/gaia}), processed by the {\it Gaia} Data Processing and Analysis Consortium (DPAC,
\url{http://www.cosmos.esa.int/web/gaia/dpac/consortium}). Funding for the DPAC has been provided by national institutions, in particular the institutions participating in the {\it Gaia} Multilateral Agreement. 

This work was (partially) supported by the Spanish MICIN/AEI/10.13039/501100011033 and by "ERDF A way of making Europe" by the European Union through grant PID2021-122842OB-C21, and the Institute of Cosmos Sciences University of Barcelona (ICCUB, Unidad de Excelencia ’Mar\'{\i}a de Maeztu’) through grant CEX2019-000918-M. FA acknowledges the grant RYC2021-031683-I funded by MCIN/AEI/10.13039/501100011033 and by the European Union's NextGenerationEU/PRTR.
In addition to those mentioned in the text, this work made use of the following software packages: \texttt{Jupyter} \citep{2007CSE.....9c..21P, kluyver2016jupyter}, \texttt{matplotlib} \citep{Hunter:2007}, \texttt{numpy} \citep{numpy}, \texttt{pandas} \citep{mckinney-proc-scipy-2010, pandas_16918803}, \texttt{python} \citep{python}, \texttt{scipy} \citep{2020SciPy-NMeth}, \texttt{astroquery} \citep{2019AJ....157...98G, astroquery_16755350}, \texttt{Cython} \citep{cython:2011}, \texttt{dustmaps} \citep{2018JOSS....3..695M, dustmaps_10517733}, \texttt{galpy} \citep{2015ApJS..216...29B}, \texttt{h5py} \citep{collette_python_hdf5_2014}, \texttt{Numba} \citep{numba:2015, Numba_10839385}, \texttt{scikit-learn} \citep{scikit-learn, sklearn_api, scikit-learn_14627164}, \texttt{seaborn} \citep{Waskom2021}, and \texttt{tqdm} \citep{tqdm_14231923}.
This research has made use of NASA's Astrophysics Data System.
Some of the results in this paper have been derived using \texttt{healpy} and the HEALPix package\footnote{\url{http://healpix.sourceforge.net}} \citep{Zonca2019, 2005ApJ...622..759G,healpy_14029333}.
Software citation information aggregated using \texttt{\href{https://www.tomwagg.com/software-citation-station/}{The Software Citation Station}} \citep{software-citation-station-paper, software-citation-station-zenodo}. This work used the accessible \texttt{Matplotlib}-like colour cycles defined in \cite{Petroff_2021}.

\end{acknowledgements}

\bibliographystyle{aa}
\bibliography{refs}

\appendix

\onecolumn
\section{Additional plots of injected clusters and cluster distribution}

\begin{figure*}[h]
\centering
\includegraphics[width=\textwidth]{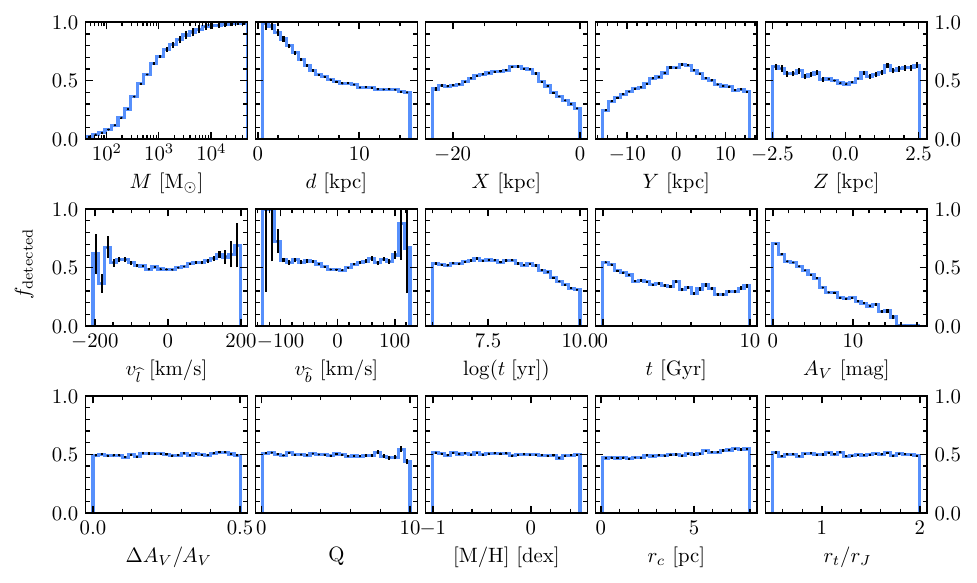}
\caption{Binned fraction of clusters detected at different parameter values for the sample of all 79\,720 injected and retrieved clusters. Poisson errors on bins are shown by the black bars. \label{fig:app:detection_trends}}
\end{figure*}

\begin{figure}[h]
\centering
\includegraphics[width=0.5\columnwidth]{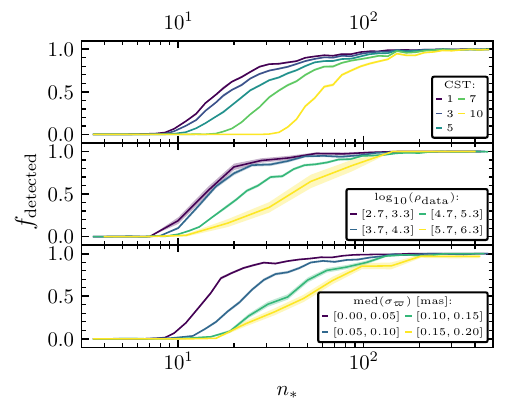}
\caption{Effect of varying cluster sample on the fraction of clusters detected against the number of injected member stars, $n_*$. Different samples are indicated by the different coloured curves, with the shaded region showing standard error. In the upper panel, CST threshold is varied. In the middle panel, only injected clusters at certain \emph{Gaia} data densities $\rho_\text{data}$ are shown. In the lower panel, clusters are divided into different samples based on the median parallax error of their simulated member stars. \label{fig:f_detected_trends}}
\end{figure}

\newpage
\section{Published data tables of simulated cluster retrievals}\label{app:published_tables}

\begin{table*}[h]
    \caption{\label{tab:clusters}Catalogue of injected and retrieved clusters.}
    \centering
    \begin{tabular}{cccccccccccccc}
        \hline\hline
        ID & Det. ID & CST & $m_\text{clSize}$ & Best? & $n_*$ & $n_\text{memb.}$ & $n_\text{noise}$ & $l$ [$^\circ$] & $b$ [$^\circ$] & $d$ [kpc] & $M$ [M\textsubscript{\sun}] & $\log t$ & $A_V$ [mag] \\
        \hline
\multicolumn{14}{c}{$\cdot \cdot \cdot$} \\

1256 & -- & 0.00 & -- & Y & 1 & 0 & 0 & 265 & 1 & 10.5 & 92 & 6.04 & 7.44 \\
1257 & sZM & 9.59 & 10 & N & 69 & 37 & 4 & 36 & -27 & 5.0 & 459 & 9.84 & 0.19 \\
1257 & AiQ & 5.71 & 10 & N & 69 & 12 & 0 & 36 & -27 & 5.0 & 459 & 9.84 & 0.19 \\
1257 & yQG & 10.64 & 20 & N & 69 & 83 & 31 & 36 & -27 & 5.0 & 459 & 9.84 & 0.19 \\
1257 & yQG & 10.64 & 40 & N & 69 & 83 & 31 & 36 & -27 & 5.0 & 459 & 9.84 & 0.19 \\
1257 & yQG & 10.64 & 80 & Y & 69 & 83 & 31 & 36 & -27 & 5.0 & 459 & 9.84 & 0.19 \\
1258 & wnZ & 18.11 & 10 & N & 187 & 152 & 7 & 76 & -27 & 0.8 & 176 & 7.65 & 0.28 \\
1258 & wnZ & 18.11 & 20 & N & 187 & 152 & 7 & 76 & -27 & 0.8 & 176 & 7.65 & 0.28 \\
1258 & wnZ & 18.11 & 40 & N & 187 & 152 & 7 & 76 & -27 & 0.8 & 176 & 7.65 & 0.28 \\
1258 & wnZ & 18.11 & 80 & Y & 187 & 152 & 7 & 76 & -27 & 0.8 & 176 & 7.65 & 0.28 \\
1259 & vVx & 2.24 & 10 & N & 22 & 24 & 12 & 225 & 0 & 10.4 & 253 & 8.51 & 2.08 \\
1259 & vVx & 2.24 & 20 & Y & 22 & 24 & 12 & 225 & 0 & 10.4 & 253 & 8.51 & 2.08 \\
1260 & -- & 0.00 & -- & Y & 2 & 0 & 0 & 43 & -1 & 10.1 & 865 & 8.00 & 9.89 \\
1261 & rsC & 5.80 & 10 & N & 38 & 37 & 13 & 204 & -2 & 14.5 & 935 & 8.19 & 2.88 \\
1261 & rsC & 5.80 & 20 & Y & 38 & 37 & 13 & 204 & -2 & 14.5 & 935 & 8.19 & 2.88 \\

\multicolumn{14}{c}{$\cdot \cdot \cdot$} \\   
        \hline
    \end{tabular}
    \tablefoot{Shown for a subset of fifteen clusters; the full version with all detections and non-detections of all simulated clusters, as well as additional columns, is available online at the CDS. Clusters may be detected at multiple $m_\text{clSize}$ values by HDBSCAN, with each detection of a cluster being assigned a unique three-character detection ID. As HDBSCAN may report identical clusters even at different $m_\text{clSize}$ values, these IDs were deduplicated -- meaning that one cluster at various $m_\text{clSize}$ values may share the same detection ID and parameters if they were identical. Clusters not detected at any $m_\text{clSize}$ value have no detection ID. The `Best?' column can be used to cut the cluster list to only the best possible detection of each cluster, which was the dataset used to train our model (in addition to later also excluding some clusters based on their $X$ coordinate; see text).}
\end{table*}

\begin{table*}[h]
    \caption{\label{tab:members}Catalogue of injected and retrieved cluster members.}
    \centering
    \begin{tabular}{cccccc}
        \hline\hline
        Simulated cluster ID & Star ID & Detection IDs & DR3 Source ID & $G$ [mag] & $G_{BP}-G_{RP}$ [mag] \\
        \hline
\multicolumn{6}{c}{$\cdot \cdot \cdot$} \\

-- & -- & yQG & 6904800569939278720 & 15.24 & 0.54 \\
-- & -- & yQG & 6904799126830387712 & 15.38 & 1.17 \\
1257 & 1887 & sZM,yQG & -- & 16.40 & 0.76 \\
1257 & 1897 & sZM,yQG & -- & 16.76 & 0.79 \\
1257 & 1856 & sZM,yQG & -- & 17.26 & 0.80 \\
1257 & 2026 & sZM,yQG & -- & 17.31 & 0.75 \\
-- & -- & yQG & 6904804379573622016 & 17.69 & 1.34 \\
1257 & 1881 & sZM,yQG & -- & 17.83 & 0.90 \\
-- & -- & yQG & 6904748274417447168 & 18.31 & 0.69 \\
1257 & 1910 & sZM,yQG & -- & 18.37 & 0.85 \\
-- & -- & yQG & 6904820356852119936 & 18.49 & 0.82 \\
1257 & 2017 & AiQ,yQG & -- & 18.50 & 0.90 \\
1257 & 1972 & sZM,yQG & -- & 18.52 & 0.88 \\
1257 & 1846 & sZM,yQG & -- & 19.37 & 1.01 \\
1257 & 1920 & -- & -- & 20.22 & 1.44 \\

\multicolumn{6}{c}{$\cdot \cdot \cdot$} \\ 
        \hline
    \end{tabular}
    \tablefoot{Shown for just fifteen assigned member stars for simulated cluster 1257 detections; the full version with all stars and all columns is available online at the CDS. To construct a membership list for a given detection, each star should be queried for if it contains the corresponding three-character detection ID. This system is used as a single star may appear in multiple different HDBSCAN detections -- including real \emph{Gaia} `noise' stars. Alternatively, a simulated cluster can be reconstructed by querying based on cluster ID, which will reproduce the original injected cluster membership.}
\end{table*}

\twocolumn
\section{Model fitting results}\label{app:model_fit_results}

\begin{table}[h]
    \caption{\label{tab:parameters}Fitted model parameters.}
    \centering
    \begin{tabular}{cccc}
        \hline\hline
        Param. & Value & Prior min & Prior max \\
        \hline

$m_n$ & $0.771^{+0.006}_{-0.011}$ & 0.0 & 10.0 \\
$b_n$ & $2.022^{+0.016}_{-0.018}$ & 2.0 & 3.0 \\
$c_n$ & $0.847^{+0.003}_{-0.005}$ & 0.0 & 5.0 \\
$d_n$ & $-1.391^{+0.012}_{-0.026}$ & -10.0 & 0.0 \\
$m_s$ & $0.119^{+0.008}_{-0.012}$ & -10.0 & 10.0 \\
$b_s$ & $1.503^{+0.001}_{-0.002}$ & 1.5 & 4.5 \\
$c_s$ & $0.012^{+0.001}_{-0.001}$ & 0.0 & 10.0 \\
$d_s$ & $-1.700^{+0.111}_{-0.358}$ & -10.0 & 10.0 \\
$m_k$ & $0.014^{+0.008}_{-0.020}$ & -10.0 & 10.0 \\
$b_k$ & $1.554^{+0.038}_{-0.036}$ & 1.5 & 4.5 \\
$c_k$ & $0.110^{+0.004}_{-0.013}$ & 0.0 & 10.0 \\
$d_k$ & $-9.913^{+0.054}_{-0.057}$ & -10.0 & 10.0 \\
$u$ & $0.006^{+0.000}_{-0.001}$ & 0.0 & 1.0 \\
$v$ & $1.678^{+0.009}_{-0.014}$ & 0.5 & 4.0 \\

        \hline
    \end{tabular}
    
\end{table}

\begin{figure}[h]
\centering
\includegraphics[width=\columnwidth]{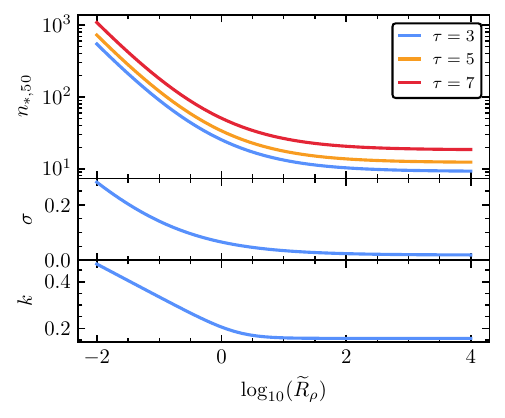}
\caption{Dependence of model parameters $n_{*,50}$, $\sigma$, and $k$ on density of a cluster $\tilde{R}_\rho$. For $n_{*,50}$, which also depends on CST threshold $\tau$, three curves are shown: one each for $\tau$ of 3, 5, and 7. \label{fig:pdet_model_params}}
\end{figure}

Table~\ref{tab:parameters} presents the results of our model fitting procedure. These parameters are used to derive $n_{*,50}$, $\sigma$, and $k$ for our logistic model, which varies with respect to $\tilde{R}_\rho$. This dependence is illustrated in Fig.~\ref{fig:pdet_model_params}, showing that the model opts for a steep, skewed logistic function at high $\tilde{R}_\rho$, reducing this to a less certain and symmetric logistic at low $\tilde{R}_\rho$. Since the median number of stars that a cluster needs to be detected in HR23 is a parameter of our model ($n_{*50}$), this can also be read directly from Fig.~\ref{fig:pdet_model_params}. Indeed, at $\tau=3$, $n_{*,50}$ is around ten, corresponding to the minimum HDBSCAN cluster size used in HR23. $n_{*,50}$ begins to rise when cluster density is about one to two orders of magnitude higher than the density of local \emph{Gaia} data. $n_{*,50}$ rises sharply at low $\tilde{R}_\rho$, with our model predicting that clusters that are 10\% as dense as \emph{Gaia} data require at least 100 stars to have a 50\% detection chance in HR23 -- a full order of magnitude higher than HR23's minimum cluster size.

\section{Advice for future all-sky cluster blind searches}\label{app:advice_for_future_catalogues}

In this section, we discuss some aspects of how HR23 and HR24 were constructed, and how they could be improved in the future to enable easier construction of cluster catalogue selection functions. These improvements could be incorporated into any future blind cluster survey, and not just those following HR23's exact methodology.

Our first difficulty arose with HR23's multiple distance bins. Clustering in HR23 was performed mostly at HEALPix level 5, with fields of nine pixels of around 30~deg$^2$ in size. However, capturing the full extent of clusters closer than around 500-750~pc required a separate run with HEALPix level 2 fields containing only stars with $\varpi<1$~mas, and capturing the full extent of clusters within around 150~pc (which are highly dispersed on-sky) required a Cartesian coordinates run on all stars within 250~pc. These three distance ranges and the 12288 separate pixels at HEALPix level 5 makes reproducing the exact method of HR23 convoluted, due to the large amount of run merging that was necessary in HR23 -- our present work hence only establishes HR23's selection function for $d>500$~pc and neglects considering the selection function of run merging (Sect.~\ref{sec:caveats}). Use of different field sizes to detect clusters of different sizes is also not unique to HR23, having also been a feature of the cluster searches of \cite{Castro-GinardJordi_2018,Castro-GinardJordi_2019,Castro-GinardJordi_2020,Castro-GinardJordi_2022}.

Originally, HR23 adopted HEALPix level 5-sized fields to save on computational time, as HDBSCAN's runtime scaling is on order $\mathcal{O}(n\log n)$ (in the best cases). This resulted in a total runtime of 8 days to construct HR23 on a modest computational node. Given the widespread use of HR23, expending more computational time to save on later hassle is probably justified. One potential easier strategy for \emph{Gaia} DR4 and beyond would be to have the main runs in a much larger HEALPix size -- such as level 2 or 3, resulting in 192 or 768 fields respectively. This would most likely still need to be aided with an additional Cartesian coordinates run to capture nearby clusters such as the Hyades without spherical distortions, but would dramatically reduce the number and complexity of merges required to create a cluster catalogue, trading computational time for ease of processing.

Our next difficulty was with HR23's sample of \emph{Gaia} data. Using a subsample of \emph{Gaia} data is a common, recommended approach that can ensure that unreliable sources are removed before analysis \citep{GaiaCollaborationVallenari_2023}. Nevertheless, the multitude of cuts applied in HR23 -- including the difficult to model machine-learning cut of \cite{RybizkiGreen_2022} -- make simulating clusters somewhat more complicated. In the future, adopting as few \emph{Gaia} data cuts as possible would likely help. If future \emph{Gaia} data releases see major improvements in the number of stars with erroneous astrometry (particularly at the faint end), then this is likely to be highly feasible.

However, in general, we have found between this work and Paper~I that establishing the selection function of a blind cluster search is eminently feasible. Our final recommendation is that cluster surveys do exactly that, using the methodology that we have developed and open-sourced in this work. The cluster census of \emph{Gaia} DR4, \emph{Gaia} DR5, and beyond should also come with an accompanying selection function that is published as soon as possible after publishing the catalogue, allowing for advanced, selection effect-corrected studies of the cluster distribution to be conducted. In the next section, we argue that this is even easier to conduct than we first anticipated.

\section{Advice for future cluster selection function studies}\label{app:advice_for_future_sfs}

\begin{figure}[t]
\centering
\includegraphics[width=\columnwidth]{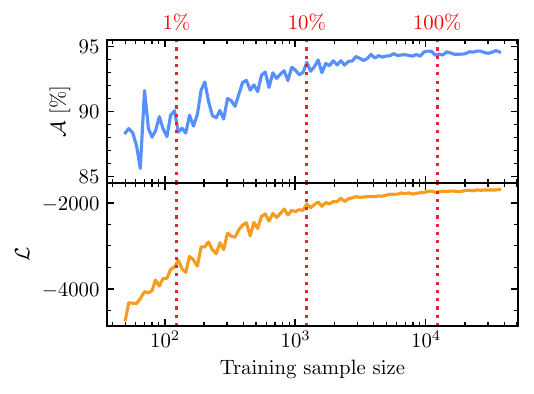}
\caption{Effect of training data sample size on the test data accuracy (\emph{upper plot}) and test data log loss (\emph{bottom plot}) of an XGBoost model trained to predict cluster detectability from $n_*$, $\tau$, $\text{med}(\sigma_\varpi)$, and $\rho_\text{data}$, as in Sect.~\ref{sec:models:xgboost}. Red dashed lines, with corresponding labels at the top of the plot, show the size of the training sample as a fraction of the total number of clusters in HR23. \label{fig:app:accuracy_train_size}}
\end{figure}

Our initial setup for this study and Paper~I was based mostly on (educated) guesses. Now, with the benefit of hindsight, we can analyse how effective certain choices were, and identify where future selection studies could be more efficient. 

The main insight we can provide is that much less computational time is necessary than we originally thought. Paper~I was based on 48\,688 clustering runs (of four clusters each) to cover the Galactic anticentre alone, which we found was more than necessary for such a small region. This work was based upon 80\,590 clustering runs (of one cluster each) to cover the entire Galaxy -- which may still have been more than necessary.

Figure~\ref{fig:app:accuracy_train_size} shows the accuracy of an XGBoost selection function predictor (as in Sect.~\ref{sec:models:xgboost}) trained on training samples of between 50 to 40\,000 clusters, applied to the same test dataset as before. Red dashed lines show the training sample size as a fraction of the original amount of computational work required to do HR23's clustering runs. Even performing around 100 clustering runs can produce a model with almost 90\% accuracy. More than 1\,000 clusters appears to have diminishing returns, with more than around 10\,000 clusters having negligible accuracy improvements. This suggests that a cluster selection function of a full-sky survey with an accuracy on par with this work could be produced by only using about the same amount of computational resources as the original survey, or even less if a few \% in accuracy is an acceptable loss. While it may still be necessary to have a larger training sample for an $n_*$ predictor as in Sect.~\ref{sec:nstar-predictor} due to its much higher number of input parameters, the comparatively smaller number of input parameters of our selection function (owing to the efficiency of $\rho_\text{data}$ and $\text{med}(\sigma_\varpi)$ in predicting detectability) means that only a surprisingly small training data set is necessary.

This could potentially be improved even further with more efficient training dataset creation. Most clusters simulated in this work are roughly in the middle of the $n_*$-$\tilde{R}_\rho$ plane in Fig.~\ref{fig:pdet_model_2d}. More efficient training data generation could sample clusters uniformly in these two parameters (instead of uniformly in parameters such as cluster mass or age), creating a smaller training dataset that more efficiently covers relevant $\tilde{R}_\rho$ and $n_*$ values required by a selection function. One such setup might simulate a large batch of clusters using the same cluster parameter distributions as in Table~\ref{tab:sampled_parameters}, also giving a large sample of simulated clusters for study; however, only a small subsample of these (which would then form a uniform $n_*$-$\tilde{R}_\rho$ distribution) would be put forward for clustering runs.

\section{Completeness and contamination of cluster memberships}\label{app:membership_analysis}

\begin{figure}[t]
\centering
\includegraphics[width=\columnwidth]{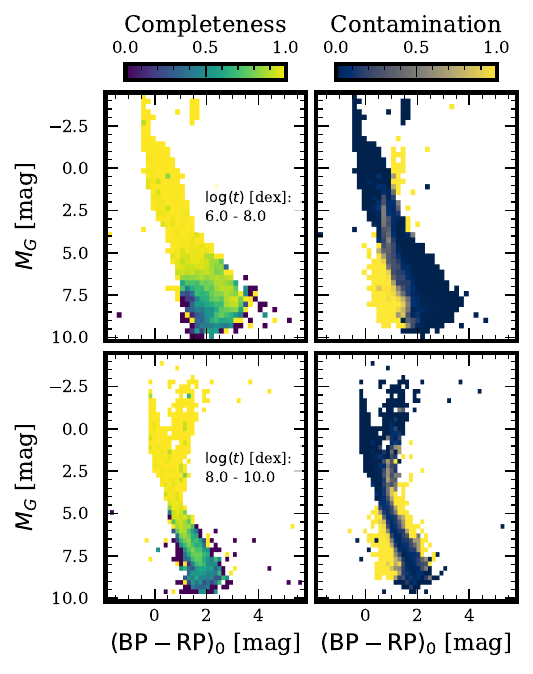}
\caption{Completeness (\textit{left}) and contamination (\textit{right}) fractions for the derived membership lists of simulated clusters shown in dereddened colour-absolute magnitude diagrams for clusters in $\log(t)$ bins of 6.0 - 8.0~dex (\textit{top}) and 8.0 - 10.0~dex (\textit{bottom}), respectively.\label{fig:memb_contamination_completeness}}
\end{figure}

While the present work aims to investigate the selection function of clusters themselves, studying the completeness and purity of cluster membership lists is equally important for a number of potential applications, such as accurately estimating cluster parameters, cluster binary fractions, and total cluster masses. Moreover, very high contamination rates can lead to some clusters being discarded by the CMD classifiers adopted by HR23 and \citet{Castro-GinardJordi_2019}, affecting the completeness of high-quality cluster samples that rely upon them.

In Fig.~\ref{fig:memb_contamination_completeness}, we show the membership-level completeness and contamination fractions for a sample of simulated clusters on a colour-absolute magnitude diagram, showing an example of how our additional published data can be used to investigate cluster membership quality. The top and bottom panels correspond to relatively young ($\log(t) \in [6.0, 8.0]$ dex) and old ($\log(t) \in [8.0, 10.0]$ dex) ranges in cluster age, respectively, with other parameters fixed for visualization purposes. As expected, in both young and old simulated clusters, the recovered upper main sequence is fairly complete, with fainter stars increasingly affected by incompleteness due to larger uncertainties in their observed astrometry. This is consistent with the findings of HR24 (see their Fig.~2), who employed a simpler method to compute the probability of inclusion of a true member in the derived membership lists of a given cluster. On the other hand, contamination by field stars affects a broader range of $G$ magnitudes, potentially biasing the estimation of cluster parameters and the detection of main sequence--main sequence and white dwarf--main sequence unresolved binaries.

Continued efforts to enhance the purity and completeness of membership lists will be essential to mitigate these effects. We hope that the simulated and recovered cluster membership lists released with this work will serve as a useful benchmark for such future improvements, or as a tool for a future work to derive direct calibration factors for HR24's membership lists.

\end{document}